\documentclass[prl,twocolumn,showpacs,amsmath,amssymb,superscriptaddress]{revtex4}

\usepackage[dvips]{graphicx}% Include figure files

\usepackage{graphics}% Include figure files

\usepackage{dcolumn}% Align table columns on decimal point

\usepackage{bm}% bold math

\usepackage[usenames]{color}

\begin{document}

\title{Tunneling anisotropic magnetoresistance in multilayer-(Co/Pt)/AlO$_{\rm x}$/Pt structures}

\author{B.~G.~Park}

\affiliation{Hitachi Cambridge Laboratory, Cambridge CB3 0HE, United Kingdom}

\author{J.~Wunderlich}

\affiliation{Hitachi Cambridge Laboratory, Cambridge CB3 0HE, United Kingdom}
\affiliation{Institute of Physics ASCR, v.v.i., Cukrovarnick\'a 10, 162 53 Praha 6, Czech Republic}

\author{D.~A.~Williams}

\affiliation{Hitachi Cambridge Laboratory, Cambridge CB3 0HE, United Kingdom}

\author{S.~J.~Joo}

\affiliation{Nano Device Research Center, Korea Institute of Science and Technology,
Seoul 136-792, Korea}

\author{K.~Y.~Jung}

\affiliation{Nano Device Research Center, Korea Institute of Science and Technology,
Seoul 136-792, Korea}

\author{K.~H.~Shin$^{\ast}$}

\affiliation{Nano Device Research Center, Korea Institute of Science and Technology,
Seoul 136-792, Korea}

\author{K.~Olejn\'{\i}k}

\affiliation{Institute of Physics ASCR, v.v.i., Cukrovarnick\'a 10, 162 53 Praha 6, Czech Republic}

\author{A.~B.~Shick}

\affiliation{Institute of Physics ASCR, v.v.i., Na Slovance 2, 182 21 Praha 8, Czech Republic}

\author{T.~Jungwirth$^{\ast}$}

\affiliation{Institute of Physics ASCR, v.v.i., Cukrovarnick\'a 10, 162 53 Praha 6, Czech Republic}

\affiliation{School of Physics and Astronomy, University of Nottingham, Nottingham NG7 2RD, United Kingdom}

\date{\today}

\begin{abstract}
We report observations of tunneling anisotropic magnetoresitance (TAMR)  in vertical tunnel devices with a ferromagnetic multilayer-(Co/Pt) electrode and a non-magnetic Pt counter-electrode separated by an AlO$_{\rm x}$ barrier. In stacks with the ferromagnetic electrode terminated by a Co film the TAMR  magnitude saturates at 0.15\% beyond which it shows only  weak dependence on the magnetic field strength, bias voltage, and temperature. For ferromagnetic electrodes terminated by two monolayers of Pt we observe order(s) of magnitude enhancement of the TAMR and a strong dependence on  field, temperature and bias. Discussion of experiments is based on relativistic {\em ab initio} calculations of magnetization orientation dependent densities of states of Co and Co/Pt model systems.
\end{abstract}


\maketitle
Anisotropic magnetoresistance (AMR) sensors replaced in the early 1990's classical magneto-inductive coils in hard-drive readheads launching the era of spintronics. Their utility has, however, remained limited partly because the response of these ferromagnetic resistors to changes in magnetization orientation originates from generically subtle spin-orbit (SO) interaction effects \cite{McGuire:1975_a}. Currently widely used giant magnetoresistance  \cite{Baibich:1988_a} and tunneling magnetoresistance (TMR) \cite{Moodera:1995_a} elements comprising (at least) two magnetically decoupled ferromagnetic layers provided a remarkably elegant way of tying the magnetoresistance response directly to the ferromagnetic exchange splitting of the carrier bands without involving SO-coupling. Large magnetoresistances in these devices are, nevertheless, obtained at the expense of a significantly increased structure complexity, necessary to guarantee independent and different magnetization switching characteristics and spin-coherence of transport between the ferromagnetic layers.

Studies of AMR  effects \cite{Gould:2004_a,Brey:2004_b,Ruester:2004_a,Wunderlich:2006_a} in ferromagnetic semiconductor tunneling devices showed that AMR response can in principle be huge and richer than TMR, with the magnitude and sign  of the magnetoresistance dependent on the magnetic field orientation and electric fields. Subsequent theoretical work predicted \cite{Shick:2006_a} that the tunneling AMR (TAMR) effect is generic in ferromagnets with SO-coupling, including the high Curie temperature transition metal systems. A detailed investigation of the TAMR is therefore motivated both by its intricate relativistic quantum transport nature and by its potential in more versatile alternatives to  current TMR devices which will not require two independently controlled ferromagnetic electrodes and spin-coherent tunneling.

Experimental demonstration of the TAMR  in a tunnel junction with a ferromagnetic metal electrode has recently been reported \cite{Moser:2006_a} in an epitaxial Fe/GaAs/Au stack. The observed TAMR  in this structure is relatively small,  bellow 0.5\%, consistent with the weak SO-coupling in Fe. In this paper we present a study of vertical tunnel devices in which the ferromagnetic electrode comprises alternating Co and Pt films. We build upon the extensive literature \cite{Celinski:1990_a,Bruno:1991_a,Ferrer:1997_a,Wilhelm:2000_a,Geissler:2001_a,Nemoto:2005_a} on ferromagnetic/heavy-element transition metal multilayers in which large and tunable magnetic anisotropies are generated at the interfaces by combined effects of induced moments and strong SO-coupling. We show that placing these interfaces adjacent to the tunnel barrier opens a viable route to highly sensitive metal TAMR structures.

\begin{figure}[h]
\vspace*{-.5cm}
\hspace*{-0cm}\includegraphics[width=.75\columnwidth,angle=90]{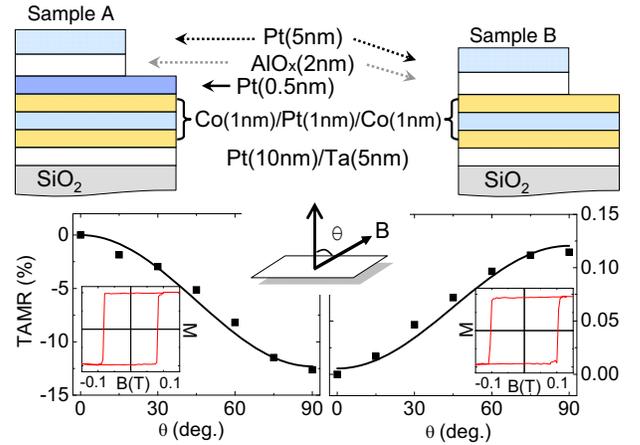}
\vspace*{-0.6cm}
\caption{Upper panels: schematic layer structures of device A (left) and device B (right). Lower panels: corresponding TAMR traces defined as $(R(\theta)-R(0))/R(0)$ at -5~mV bias and 4~K. Insets: SQUID magnetization measurements in out-of-plane magnetic fields.}
\label{f1}
\end{figure}

Our tunneling devices, schematically illustrated in Fig.~1, were grown by magnetron sputtering on a thermally oxidized Si wafer. The Ta/Pt seed layer was chosen to initiate  growth of a textured Pt(111)/Co ferromagnetic film with a strong out-of-plane magnetocrystalline anisotropy. The tunnel barrier was fabricated by plasma oxidation of  1.6~nm Al layer and the growth was completed by sputtering a non-magnetic Pt counter-electrode. Two types of multilayers are investigated: samples of type A have the alternating sequence of Pt and Co layers beneath the AlO$_{\rm x}$ barrier terminated by  0.5~nm (two-monolayer) Pt film while samples B have this top Pt film in the ferromagnetic electrode omitted. Pillar transport devices of 10-100~$\mu$m diameter were patterned in the multilayers using photolithography and ion etching with corresponding  resistances of the order $\sim 10-100$ k$\Omega$. Devices were mounted on a rotating sample holder allowing the application of magnetic fields up to 10~T at an arbitrary in-plane and out-of-plane angle. Four-point transport measurements were used to determine the TAMR.
%An additional in-plane anisotropy was introduced by applying a constant in-plane magnetic field of the order of 0.1T during the film growth along one of the cleave orientations of the sample.
\begin{figure}[h]
\vspace*{-.2cm}
\hspace*{-0cm}\includegraphics[width=.75\columnwidth,angle=90]{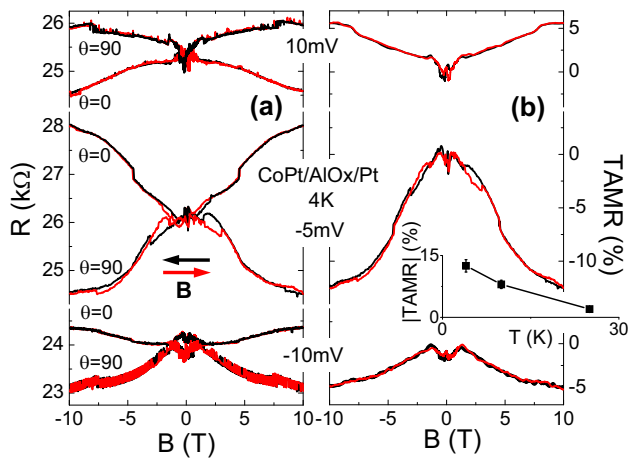}
\vspace*{-0.5cm}
\caption{(a) Resistance and (b) TAMR defined as $(R(90^{\circ})-R(0))/R(0)$ (b) of sample A measured at 10, -5, and -10~mV bias and at 4~K. Inset: temperature dependence of the TAMR for the -5~mV bias and 10~T field.}
\label{f2}
\end{figure}

Lower panels of Fig.~1 compare characteristic magnetization and magnetotransport anisotropy data measured in  samples A and B.  Wide and square hysteresis loops in perpendicular magnetic fields, shown in the insets, confirm  the presence of strong magnetic anisotropy with out-of-plane easy axis in both devices. The anisotropic magnetotransport in the two tunneling devices is, however, fundamentally different. The TAMR traces plotted in  Fig.~1 are defined as $(R(\theta)-R(0))/R(0)$ and are taken at 10~T magnetic field rotating from the perpendicular ($\theta=0$) to the in-plane ($\theta=90^{\circ}$) direction. In both samples the TAMR has the expected uniaxial symmetry but the magnitude of the effect is enhanced by 2 orders of magnitude in sample A. As the transport characteristics of TMR or TAMR tunneling devices are expected to be strongly influenced by the nature of the surface layers of the ferromagnetic electrode \cite{Tsymbal:1997_a,Chantis:2006_a} we attribute the large TAMR signal in sample A to the induced moment and strong SO-coupling in the two-monolayer Pt film inserted between Co and the tunnel barrier. In sample B, the nearest Pt/Co interfaces to the barrier is covered by 1~nm (five-monolayer) Co film and therefore this tunneling device behaves effectively as a Co/barrier/normal-metal stack for which the magnetotransport anisotropy is expected to be comparably weak as in the previously studied Fe-based TAMR device \cite{Shick:2006_a,Moser:2006_a}.

The distinct phenomenologies of samples A and B are further detailed in Figs.~2 and 3. Fig.~2(a) shows field-sweep magnetoresistance measurements in sample A for $\theta=0$ and $90^{\circ}$ at three different bias voltages and the corresponding TAMR curves are plotted in Fig.~2(b). The initial onset of the TAMR, seen clearly in the 10~mV bias measurement and associated with rotation of the magnetization from the out-of-plane easy-axis direction in the in-plane field sweep, is followed by a further dependence of the TAMR  on the field strength at higher fields with no signs of saturation  up to the maximum measured field of 10~T. Since appreciable magnetoresistance is observed in both in-plane and out-of-plane field sweeps and the sign of the magnetoresistance changes from positive to negative for different bias voltages we exclude the Lorentz force origin of the TAMR in our thin tunnel-barrier devices. Apart from the strong field and bias dependence we also observe large variation of the TAMR magnitude with temperature (see inset in Fig.~2). These observations are unique to the samples of type A; samples of type B show only weak dependence of their TAMR on field strength, bias and temperature, as shown in Fig.~3.

\begin{figure}[h]
\vspace*{-1.5cm}
\hspace*{-0cm}\includegraphics[width=.75\columnwidth,angle=90]{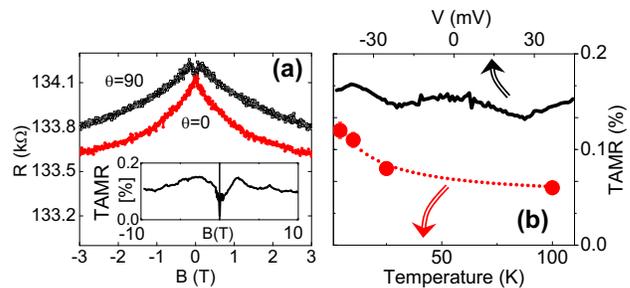}
\vspace*{-1.5cm}
\caption{(a) Resistance and TAMR (inset) defined as $(R(90^{\circ})-R(0)/R(0)$ of sample B at 10~mV bias and 4~K. (b) Temperature and bias dependence of the TAMR at 10~T field.}
\label{f3}
\end{figure}
Our theoretical discussion of the observed TAMR phenomenologies in the two types of Co/Pt tunneling devices  follows the approach applied previously to ferromagnetic semiconductor TAMR structures \cite{Gould:2004_a}. The analysis is based on calculating density of states (DOS) anisotropies in the ferromagnetic electrode with respect to the orientation of the magnetic moment and assumes proportionality between the DOS and the differential tunneling conductance anisotropies.

The SO-coupled band structures are obtained within the local spin density approximation using the relativistic version of the full-potential linearized augmented plane-wave method (FP-LAPW) in which SO interaction is included in a self-consistent second-variational procedure \cite{Shick:1997_a,Shick:2006_a}. The magnetic force theorem \cite{Liechtenstein:1987_a} is used to evaluate the DOS anisotropy: starting from self-consistent charge and spin densities calculated for the magnetic moment aligned along one of principal axes, the moment is rotated and a single energy-band calculation is performed for the new orientation of magnetization. The DOS anisotropies result from SO-coupling induced changes in the band eigenvalues. In order to increase the accuracy in DOS evaluation, the smooth Fourier interpolation scheme of Pickett {\em et al.} \cite{Pickett:1988_a} is used together with the linear tetrahedron method.

To model sample A we performed thin-slab calculations for five-monolayer Co sandwitched between two-monolayer Pt films. Sample B is modeled by considering the thin Co film only. The calculated DOS as a function of energy measured from the Fermi level are shown in Fig.~4(a) for the in-plane and out-of-plane magnetization orientations. For the Co/Pt model system we find a complex structure of the DOS near the Fermi energy with the main features shifted by 10's of meV when comparing  the two magnetization orientation curves. The Co-film, on the other hand, has a nearly featureless DOS near the Fermi level which depends very weakly on the magnetization direction. The relative difference between DOSs for the in-plane and out-of-plane magnetizations, which we relate to the tunneling conductance anisotropy, shows an oscillatory behavior as a function of energy for the Co/Pt slab with a magnitude of up to $\sim 20$\%. For the Co film the magnitude and the energy dependence of the anisotropy is substantially weaker. These theoretical results are qualitatively consistent with measurements in samples A and B presented above.

To facilitate a more direct comparison between theory and experiment we show in the lower panels of Fig.~4 the measured differential conductances and the corresponding conductance anisotropies as a function of the bias voltage in sample A. Both the relative shift of the differential conductances for the two magnetization orientations and the oscillatory behavior of the conductance anisotropy with changing bias reflect qualitatively the  anisotropic DOS behavior of the Co/Pt system seen in the theory curves.

In the calculations shown in Fig.~4 we assumed relaxed Co-Co and Co-Pt interlayer spacings in the stack. To estimate effects of lattice parameter variations on the TAMR we performed additional calculations (see Fig.~5(a)) with the Co-Co spacing in the film corresponding to bulk hcp Co. Qualitatively we obtained similar shifts in the complex DOS patterns for the two magnetization orientations and comparable overall magnitude of the DOS anisotropies but the details of the oscillatory dependence on the energy are markedly different from the calculations in Fig.~4.
\begin{figure}[h]
\vspace*{-0.3cm}
\hspace*{-0cm}\includegraphics[width=.8\columnwidth,angle=90]{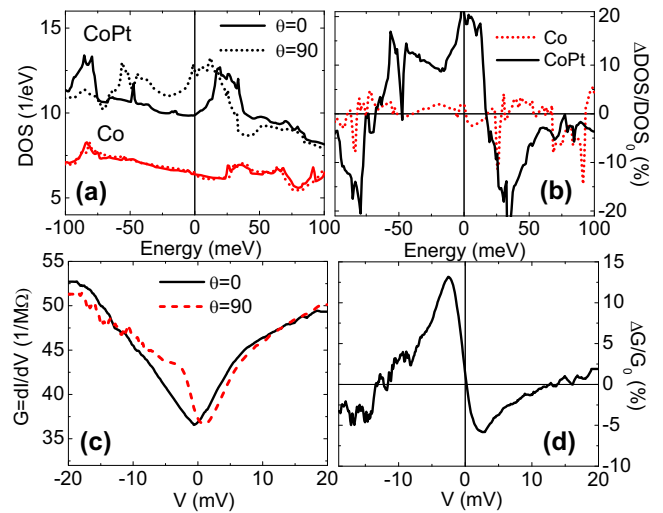}
\vspace*{-0.5cm}
\caption{(a) Theoretical DOSs at $\theta=0$ and $90^{\circ}$ and (b) the relative DOS anisotropy defined as $(DOS(90^{\circ})-DOS(0)/DOS(0)$ as a function of energy measured form the Fermi level for the Pt/Co and Co model systems. (c) Corresponding experimental differential conductances and (d) relative differential conductance anisotropies as a function of bias.}
\label{f4}
\end{figure}

Theoretical expectation that lattice distortions can significantly alter the TAMR response is echoed in complementary experiments, shown in Fig.~5(b), where we study the TAMR as a function of the in-plane angle in samples of type A and B. A uniaxial in-plane magnetocrystalline anisotropy is generated in these samples by applying in-plane magnetic field during the growth. This is an established procedure to control magnetic anisotropies in transition metal structures and in Co/Pt systems has been ascribed to anisotropic bonding effects at the Co/Pt interface \cite{Einax:2007_a}. We can therefore expect that the induced in-plane magnetic anisotropy will be reflected in the TAMR characteristics only for sample A in which the Pt/Co interface is in contact with the tunnel barrier. Indeed Fig.~4 shows a negligible dependence in sample B of the TAMR on the in-plane angle $\varphi$ towards which magnetization is rotated from the out-of-plane direction. In sample A, on the other hand, the dependence on $\varphi$ has a uniaxial character consistent with the induced in-plane magnetocrystalline anisotropy and, notably, the magnitude of the variation with  $\varphi$ is comparable to the magnitude of the $\theta$-dependent TAMR shown in Fig.~1. This result illustrates the potential for controlling the characteristics of highly sensitive TAMR devices by lattice structure engineering. We also note that the observation of the large in-plane TAMR signal in our devices confirms the SO-coupling origin of the measured tunneling anisotropy effects.
\begin{figure}[h]
\vspace*{-0cm}
\hspace*{-0cm}\includegraphics[width=.8\columnwidth,angle=90]{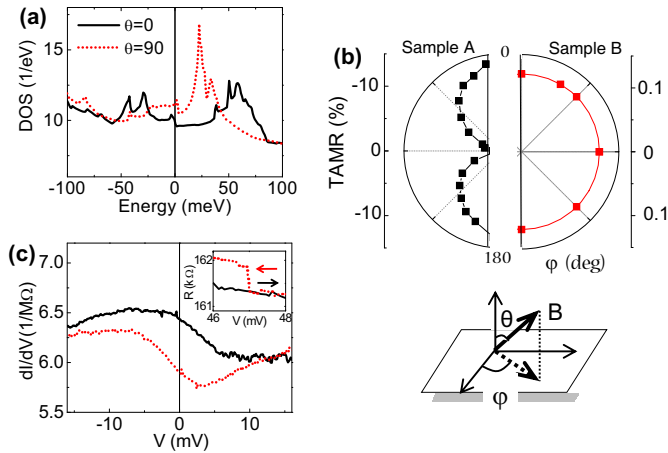}
\vspace*{-1cm}
\caption{(a) Theoretical DOSs at $\theta=0$ and $90^{\circ}$ for an unrelaxed Pt/Co model systems. (b) Experimental TAMR defined as $(R(\theta=90^{\circ},\varphi)-R(\theta=0)/R(\theta=0)$ for samples A and B.  (c) Resistance measurements (inset) in two independent bias sweeps showing a discrete step in resistance for the upper curve. Corresponding differential conductances (main panel) showing a relative shift in the oscillatory bias dependence for the two measurements.}
\label{f5}
\end{figure}

We conclude with a brief discussion of the observed magnetic field and temperature dependence of the TAMR in sample A and implications our work suggests towards the realization of TAMR structures with optimized sensitivity and field and thermal stability. Enhanced, compared to bulk systems, and temperature dependent susceptibility of heavy element transition metals has been studied in thin films exchange coupled to neighboring ferromagnetic layers or in dilute ferromagnetic alloys \cite{Crangle:1965_a,Celinski:1990_a}. Depending on the thickness of the film or concentration of the ferromagnetic elements in the dilute alloys the systems can exhibit superparamagnetic behavior ascribed to the presence of fluctuating moments in the heavy element transition metals \cite{Moriya:1985_a}. Experimentally, magnetization in these systems can show Curie-Weiss-like scaling with $B/T$, and remain unsaturated at high fields and low temperatures. We infer that similar physics applies to our two-monolayer Pt film coupled to the neighboring Co layer or to intermixed Co atoms in the Pt. The observed approximate scaling of the TAMR with $B/T$, implicitly contained in Fig.~2, can then be partly ascribed to the proportionality between TAMR and the unsaturated polarization of the Pt layer adjacent to the tunnel barrier.

Another effect which contributes to the suppression of the TAMR in our devices with temperature is telegraphic noise associated with randomly trapped charges presumably in the AlO$_{\rm x}$ barrier or at the interface between the barrier an the Pt thin film \cite{Nowak:1999_a}. At low temperatures, the resulting jumps between distinct resistance states can be clearly identified, as shown in Fig.~5(c), together with shifts in the corresponding oscillatory bias dependence of the differential conductance. These jumps become more frequent with increasing temperature which leads to averaging out of the TAMR signal.

To summarize, our work demonstrates the prospect for realizing highly sensitive TAMR devices in  transition metal structures. Strategies for future research of these systems should include fine tuning of the choice of the transition metals  and of the thickness and composition of the films to optimize the SO-coupling and exchange splitting of layers adjacent to the tunnel barrier. Improving the crystalline quality of the barrier and the barrier/electrode interfaces should also lead to enhanced magnitude and stability of the TAMR effect.

We acknowledge  support from  the TND Frontier Project funded by KISTEP,
the KIST Institutional program, from the Korea Research Council of
Fundamental Science \& Technology (KRCF), from the EU Grant  IST-015728, UK Grant GR/S81407/01, and Czech Republic  Grants 202/05/0575, 202/04/1519, FON/06/E002, AV0Z1010052, and LC510. $^{\ast}$ Corresponding author.

%\bibliography{MSWEBpublications}

\end{document}